\documentclass[twocolumn]{openjournal}
\usepackage{siunitx}
\usepackage{graphicx} % Required for inserting images
\usepackage[colorlinks,linkcolor=blue,citecolor=blue,urlcolor=blue ]{hyperref}
\usepackage{listings}
\usepackage{natbib}

\begin{document}

% \maketitle
\title{Semantic Segmentation of Solar Radio Spikes at Low Frequencies}

\author{Pearse C. Murphy$^\star$,}
\thanks{$^\star$\href{mailto:pearse.murphy@obspm.fr}{pearse.murphy@obspm.fr}}
\affiliation{LESIA, Observatoire de Paris,  Université PSL, Sorbonne Université, Université Paris Cité, CNRS, 92190  Meudon, France}
\affiliation{Physics, School of Natural Sciences, University of Galway, H91 TK33, Galway, Ireland}
\author{Stéphane Aicardi}
\affiliation{DIO, Observatoire de Paris, Université PSL, CNRS, 75014 Paris, France}
\author{Baptiste Cecconi}
\affiliation{LESIA, Observatoire de Paris,  Université PSL, Sorbonne Université, Université Paris Cité, CNRS, 92190  Meudon, France}
\author{Carine Briand}
\affiliation{LESIA, Observatoire de Paris,  Université PSL, Sorbonne Université, Université Paris Cité, CNRS, 92190  Meudon, France}
\author{Thibault Peccoux}
\affiliation{LESIA, Observatoire de Paris,  Université PSL, Sorbonne Université, Université Paris Cité, CNRS, 92190  Meudon, France}
\affiliation{IPSA, 94200 Ivry-sur-Seine, France}

\begin{abstract}
    Solar radio spikes are short lived, narrow bandwidth features in low frequency solar radio observations.
    The timing of their occurrence and the number of spikes in a given observation is often unpredictable.
    The high temporal and frequency resolution of modern radio telescopes such as NenuFAR mean that manually identifying radio spikes is an arduous task.
    Machine learning approaches to data exploration in solar radio data is on the rise.
    Here we describe a convolutional neural network to identify the per pixel location of radio spikes as well as determine some simple characteristics of duration, spectral width and drift rate.
    The model, which we call SpikeNet, was trained using an Nvidia Tesla T4 \qty{16}{\giga\byte} GPU with \num{\sim 100000} sample spikes in a total time of 2.2 hours.
    The segmentation performs well with an intersection over union in the test set of $\sim 0.85$.
    The root mean squared error for predicted spike properties is of the order of $23\%$.
    Applying the algorithm to unlabelled data successfully generates segmentation masks although the accuracy of the predicted properties is less reliable, particularly when more than one spike is present in the same $64 \times 64$ pixel time-frequency range.
    We have successfully demonstrated that our convolutional neural network can locate and characterise solar radio spikes in a number of seconds compared to the weeks it would take for manual identification.
    
\end{abstract}
\maketitle

\section{Introduction}
\label{intro}

Solar radio bursts occur over a wide frequency range with varying spectral and temporal characteristics.
At decametric wavelengths there are five canonical types of radio burst which can have durations ranging from seconds to hours and spectral widths up to tens of \unit{\mega\hertz}.
Alongside these, there exist a wide variety of short duration, narrow spectral width bursts.
These bursts generally fall into three categories; solar radio spikes \citep{tarnstrom_solar_1972}, S-bursts \citep{Ellis1969} and striae \citep{DeLaNoe1972}, with individual striae and spikes sharing a number of characteristics \citep{clarkson_first_2021}.
These short lived bursts can give an insight into energy transfer that takes place over a number of milliseconds \citep[e.g.][]{clarkson_first_2021} as well as the turbulent nature of plasma in the corona \citep[e.g.][]{Kontar2017, reid_fine_2021}.
They may also offer a way of remotely measuring the magnetic field in the corona \citep[e.g.][]{Melnik2010, Clarke2019}.
This underscores the importance of understanding the origins of short duration, narrow bandwidth solar radio bursts.

Modern state of the art radio telescopes such as the LOw Frequency ARray \citep[LOFAR;][]{VanHaarlem2013} and its extension, the New Extension in Nançay Upgrading LOFAR \citep[NenuFAR;][]{zarka_lssnenufar_2012} have the temporal and spectral resolution necessary to analyse short duration, narrow bandwidth solar radio bursts in great detail \citep[][Briand et al. \textit{in prep}]{Morosan2015, Clarke2019, Kontar2017, Sharykin2018, clarkson_first_2021, clarkson_solar_2023}.
However, the sporadic occurrence of spike bursts coupled with the vast data rates of the high resolution observations means that finding them manually can be an arduous and time consuming task.
Not only this but manual identification of spikes can be biased towards the largest and brightest spikes which thus biases the time and frequency distribution of events.
Modern machine learning methods are the ideal solution for this as, once the model is trained, they eliminate the need for human identification of spikes and determining their characteristics.
The application of machine learning algorithms to solar radio data has seen increased popularity in recent literature.
Already, we have seen the application of the YOLO \citep[you only look once;][]{redmon_you_2016} convolutional neural network (CNN)  to detecting solar radio spikes in the \qtyrange[range-phrase=--]{1.1}{1.34}{\giga\hertz} range \citep{lv_automated_2023}, while \cite{Scully2021, scully_improved_2023} used YOLO to detect Type III radio bursts in the \qtyrange[range-phrase=--]{10}{90}{\mega\hertz} range.

YOLO performs what is known as object detection, where an object is located in an image and a bounding box which determines its height and width is drawn around it.
The task of locating an object in an image on a per pixel basis is known as semantic segmentation.
The varied and irregular morphology of solar radio spikes make it a problem particularly suited for the use of semantic segmentation.
\cite{murphy_automatic_2024} used a CNN called UNET \citep{ronneberger2015} to perform semantic segmentation on NenuFAR dynamic spectra to automatically detect the presence of solar radio emission.
Here we adapt the work of \cite{murphy_automatic_2024} to perform segmentation for solar radio spikes and determine their location, duration, spectral width and drift rate in NenuFAR dynamic spectra.

The design and training procedure for our model is described in Section \ref{method}.
We describe the results of the training in Section \ref{results} and give further discussion on the application of our CNN in Section \ref{discussion}. 
We conclude with a summary in Section \ref{conclusion}.

\section{Method}
\label{method}

Our goal is to develop a CNN to automatically generate a mask of where a solar radio spike occurred in the time-frequency domain as well as predict the spike's central location, duration, spectral width and drift rate.
In order to achieve this, we must first perform pre-processing of the NenuFAR dynamic spectra in order to create a dataset suitable for training the CNN.

\subsection{Preparing the datasets}
\label{data_prep}

\begin{figure}
    \centering
    \includegraphics[width=\columnwidth]{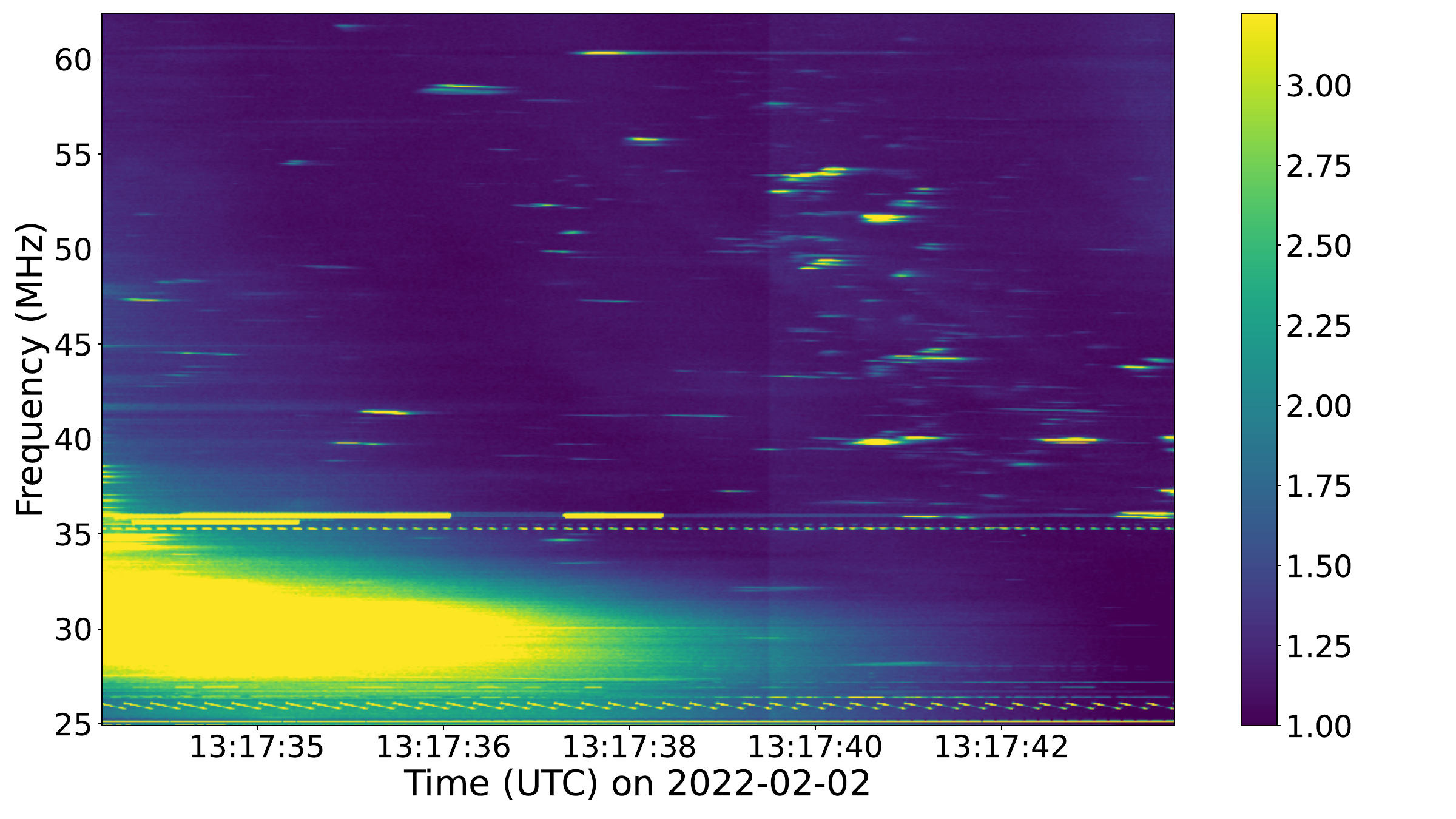}
    \caption{A sample of solar radio spikes on 2022-02-02. The colourbar is intensity relative to the background. The large number of spikes in such a short time range indicate the difficulty in manually annotating their every occurrence. Radio frequency interference and the end of a type III solar radio burst in the frequency range \qtyrange[range-phrase=--]{25}{35}{MHz} also pose a challenge.}
    \label{fig:spike-sample}
\end{figure}

The time and frequency location, spectral width, duration and drift rate of \num{1000} spike bursts were identified manually.
The spikes were located between \qty{20}{\mega\hertz} and \qty{85}{\mega\hertz} in NenuFAR observations on 2022-02-02, 2022-05-29, 2023-05-02, 2023-06-01 and 2023-07-10.
These are the same data used by Briand et al. \textit{(in prep)}.
Figure \ref{fig:spike-sample} shows a dynamic spectrum of a number of radio spikes occurring on 2022-02-02 over \qty{\sim 9}{\second} in the frequency range \qtyrange[range-phrase=--]{25}{60}{\mega\hertz}. The figure also shows radio frequency interference (RFI) which is a common feature of low frequency observations and thus our dataset must incorporate this.

These spikes formed the basis for the training, validation and test datasets. We performed a stratified split based on the date of observation so that each split had the same proportion of spikes that were identified on each date.
The percentage of spikes in each split were 64\% training, 16\% validation and 20\% test.
The inputs to the CNN are $64 \times 64$ pixel Stokes I and Stokes V/I dynamic spectra which we hereafter refer to interchangeably as tiles.
We chose $64 \times 64$ in an effort to capture the full time and spectral extent of an individual spike burst and to minimise overlap with nearby spikes.
This tile size also avoids ``out-of-memory" issues that occurred when attempting to train on larger sizes.
One time pixel corresponds to \qty{\sim 21}{\milli\second} and one frequency pixel is \qty{\sim 6.1}{\kilo\hertz}.
The time and frequency were obtained from channelisation of the native NenuFAR resolution of \qty{\sim 5.12}{\micro\second} and \qty{\sim 195.3125}{\kilo\hertz} using the Unified Dynamic Spectrum Pulsar and Time Domain receiver \citep[UnDySPuTeD;][]{Bondonneau2020a}.
In order to have a uniform time and frequency resolution for all observations, further down sampling to the desired \qty{\sim 21}{\milli\second}, \qty{\sim 6.1}{\kilo\hertz} resolution was performed using the \texttt{nenupy} Python library \citep{alan_loh_2020}.
This results in a time duration of \qty{1.34}{\second} and a frequency range of \qty{390.625}{\kilo\hertz} for each tile.
CNNs often see improved performance when their input values lie within a small range, thus we normalised the intensities to the background and rescaled in the range \numrange[range-phrase=--]{0}{1}.
The frequency range of the dynamic spectrum in \unit{\mega\hertz} was also included as an input.
In order to generate the ground truth output for the model, a segmentation mask was produced for each spike burst using the CUSUM-slope method described by \cite{murphy_automatic_2024}. 

To increase the number of samples available for training, we applied random shifts in time and frequency to the spikes in the training and validation sets.
These shifts were applied by finding the centre of the tile and adding a time and frequency offset, the magnitude and direction of which were randomly determined.
The maximum possible shift in time was the duration of the spike while the maximum possible shift in frequency was \qty{195.3125}{\kilo\hertz}.
A new $64 \times 64$ pixel tile was then generated using this offset point as its centre.
The same shifts were then applied to the corresponding segmentation masks.
In an effort to improve the robustness of the model when dealing with unseen data, we also located times where no spike occurred to produce samples of the background and added these to the training set. These also included samples in common RFI bands $\lesssim$ \qty{30}{\mega\hertz} and at \qty{\sim 70}{\mega\hertz}.
In total the training set comprised of \num{90435} samples and the validation set \num{16000}.
The test set did not undergo any augmentation and contained 200 samples.

Each spike in the dataset was manually identified and labelled so that we have a list of its characteristics, i.e., its location in time and frequency, its duration, its spectral width and its drift rate.
Full details for the manual identification will be given in Briand et al. (\textit{in prep}) and we briefly list them here.
The central time and frequency are determined by the time/frequency of the maximum stokes I intensity of the spike, $I_{max}$.
The duration and spectral width are the time/frequency extent at 20\% above the background intensity in the temporal/spectral direction from $I_{max}$.
The drift rate is calculated by applying a linear fit to the points of maximum intensity along the spectral width of the spike.
We determine the time and frequency location in terms of pixels while duration, spectral width and drift rate are in physical units of \unit{\second}, \unit{\mega\hertz} and \unit{\mega\hertz\per\second} respectively.

\begin{figure}
    \centering
    \includegraphics[width=\columnwidth]{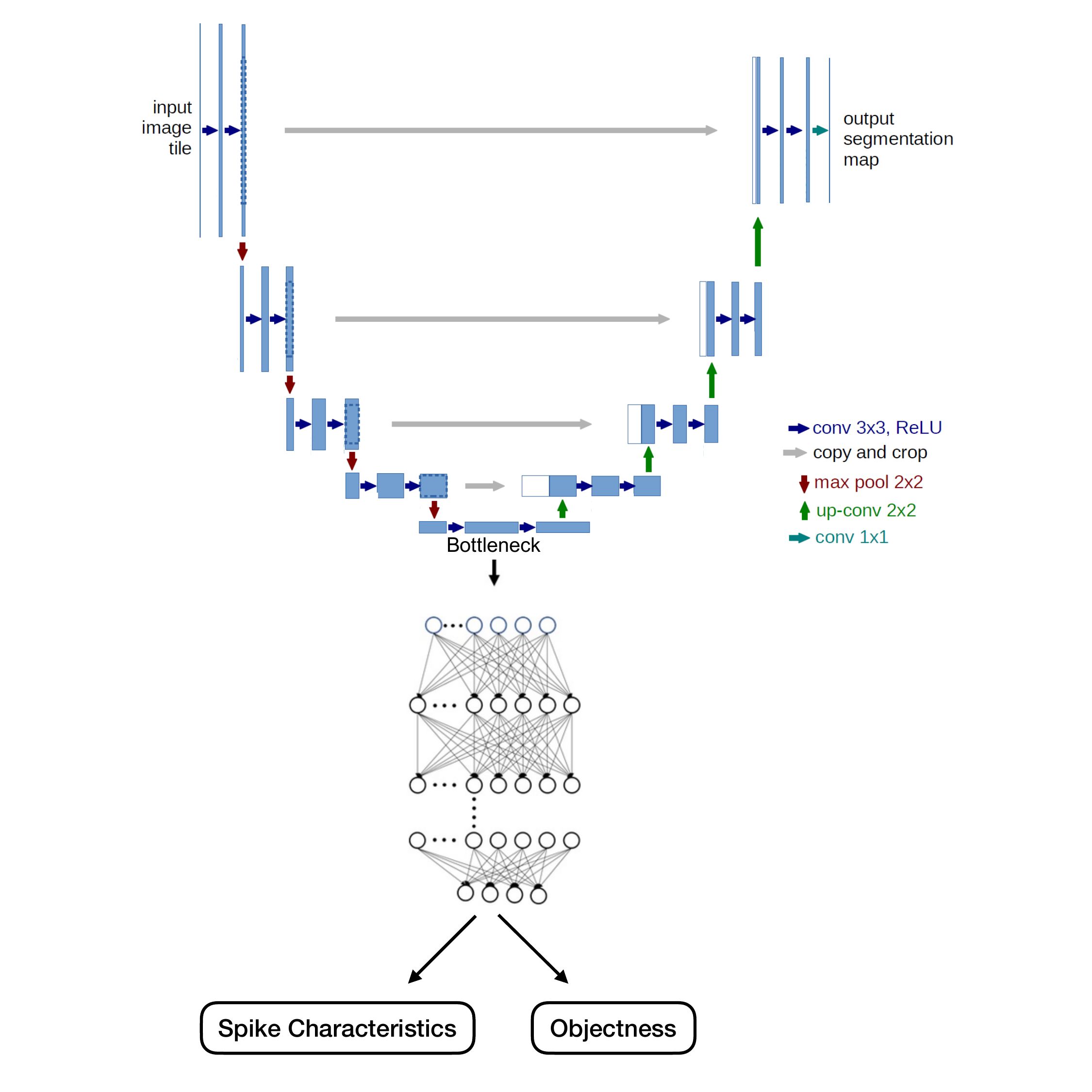}
    \caption{Architecture of our model. The top half of the model is the UNET adapted from \cite{ronneberger2015}. We pass the outputs from UNET's bottleneck layer to a fully connected network. This in turn has two outputs; the spike characteristics and the objectness.}
    \label{fig:model_arch}
\end{figure}

Finally, we determined an objectness score for each sample.
This is used to determine how certain the model is that a spike exists in the sample at all.
An objectness of 0 means the model is certain there is no spike while an objectness of 1 means the model is certain there is a spike.
We determine the objectness of each sample as the ratio of total number of pixels in the shifted segmentation mask to that of the segmentation mask before it was shifted.
Thus, any spike that is not fully in a tile will have an objectness score less than one.

\subsection{Model architecture and training the model}

The primary output of our model, affectionately referred to as SpikeNet, is a mask of every pixel where a solar radio spike occurred in time and frequency.
As mentioned in Section \ref{intro}, this type of task is known as semantic segmentation and a common CNN used in this regard is called UNET.
We use the implementation of UNET in the \texttt{segmentation\_models} Python library \citep{Yakubovskiy:2019} as a starting point.

As well as trying to produce a segmentation mask from these inputs, we require our model to determine the central location in time and frequency, duration, spectral width and drift rate of the spike.
Thus, in addition to the decoder part of the UNET, we included a fully connected deep neural network with 5 hidden layers of 30 neurons each to determine these characteristics, as well as the objectness score.
The model architecture is depicted in Figure \ref{fig:model_arch}. We take the values in the bottleneck layer of the original UNET to use as the input to the fully connected network. 
The final hidden layer is connected to two separate output layers resulting in two separate outputs, the 5 spike characteristics and the objectness score.

Training a machine learning model such as SpikeNet involves computing the difference between the model's predictions and the ground truth in what is known as a loss function. During training the weights and biases in the network are updated in order to minimise this loss function.
We apply a different loss for each of the outputs; binary cross entropy for the segmentation performed by UNET, mean squared error for the regression of spike characteristics and binary cross entropy for the objectness.
Binary cross entropy is typically used for tasks that look to determine the probability of an instance belonging to one of two classes, i.e. classification.
Mean squared error, on the other hand, is more appropriate for tasks that compare how close a predicted value is to the true value, i.e. regression.
The total loss measured during training was the weighted sum of each of these losses.
We give a weight of 10 to the regression loss and a weight of 1 to segmentation and objectness loss in  order to prioritise performance in the regression task.

Before SpikeNet undergoes any training, its predictions will be far from the truth which can lead to erratic behaviour in the early stages of training.
To combat this, we apply a warm-up period for the first 5 epochs of training by linearly increasing the training rate.
If the training rate is too large, it will take longer to converge on a global minimum in the loss function.
We thus implement a cosine decay for the training rate for the remainder of training. %as shown in Figure \ref{fig:lratevsepoch}.
The total validation loss was monitored per epoch and the training was stopped once it failed to decrease for three epochs in a row.
The model was trained for a total of 26 epochs which took 2.2 hours on an Nvidia Tesla T4 \qty{16}{\giga\byte} GPU.

% \begin{figure}
%     \centering
%     \includegraphics[width=\columnwidth]{spike_training_lrate_batch_size_32_regularisation_EarlyStopping_hidden_layers_5_neurons_per_layer_30_n_params_5_20240118.png}
%     \caption{Learning rate for each epoch. A warmup phase of 5 epochs is followed by a cosine decay before early stopping at epoch 25.}
%     \label{fig:lratevsepoch}
% \end{figure}
\begin{figure}[t]
    \centering
    \includegraphics[width=\columnwidth]{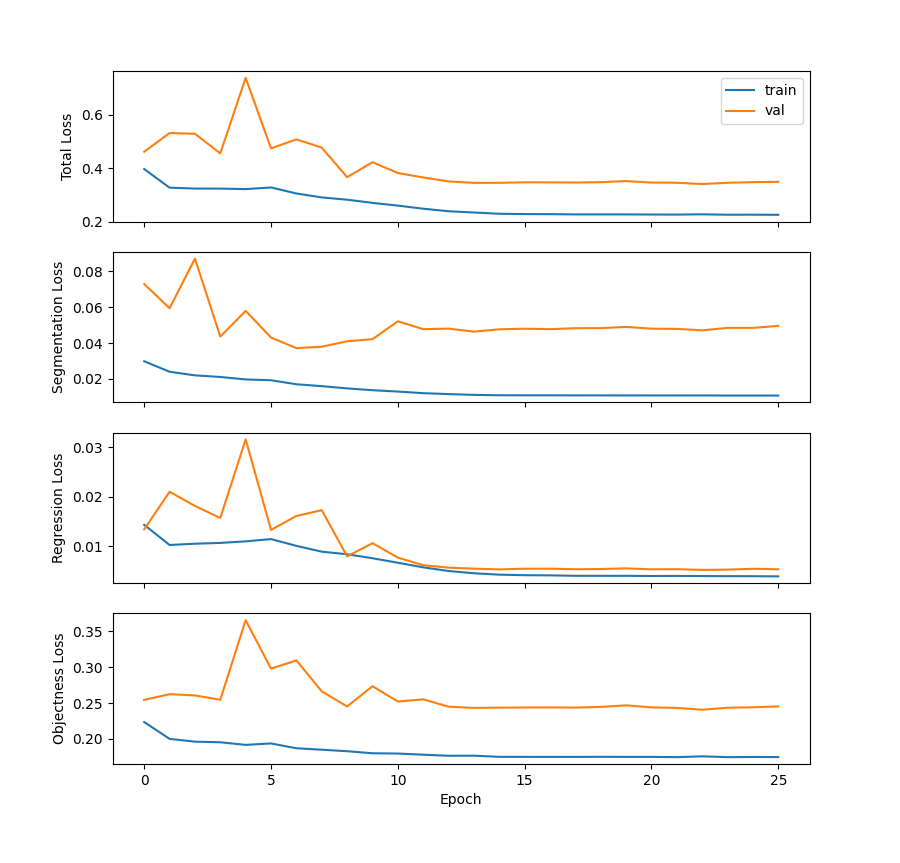}
    \caption{Evolution of loss with respect to training epoch. The training (blue) and validation (orange) loss is shown in each panel for the total loss, segmentation loss, regression loss and objectness loss.}
    \label{fig:lossvsepoch}
\end{figure}
\section{Results}
\label{results}

The values of the individual and total loss functions per epoch are shown in Figure \ref{fig:lossvsepoch}.
The loss for the training set (blue) decreased rapidly in all cases while the validation loss (orange) decreased slowly before forming a plateau. 
The gap between the training and validation loss could indicate under fitting or that the data samples in the training set were not representative of those in the validation set.

\begin{figure}
    \centering
    \includegraphics[width=\columnwidth]{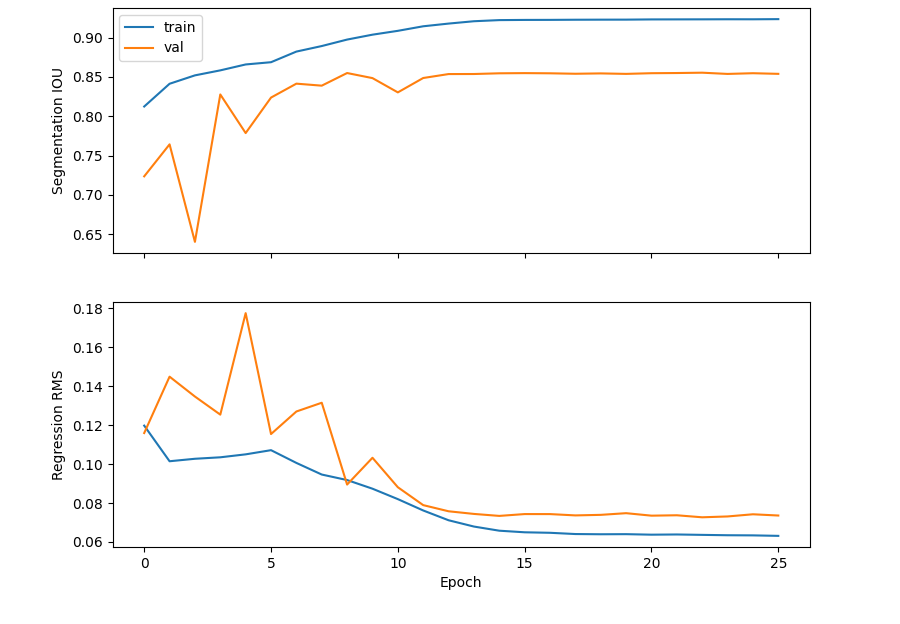}
    \caption{The evolution of model metrics with increasing training epoch. The top panel shows the intersection over union for the training (blue) and validation (orange) sets. Similarly, the root mean squared error is shown in the bottom panel.}
    \label{fig:metricvsepoch}
\end{figure}

The key metric to determine the accuracy of our segmentation masks to the ground truth is the intersection over union (IoU) or Jaccard index.
It is defined, as the name suggests, as
\begin{equation}
    \text{IoU}(A,B) = \frac{|A \cap B|}{|A \cup B|},
\end{equation}
where $A$ is the ground truth and $B$ is the prediction.
The top panel of Figure \ref{fig:metricvsepoch} shows the IoU for the validation set reaching 
$\sim 85\%$.
The bottom panel of Figure \ref{fig:metricvsepoch} shows that the root mean squared (RMS) error of the regression to predict spike characteristics decreases with each epoch. Comparing the mean RMS for all characteristics of 0.0736 to the mean of the true characteristics gives an approximate error of $23\%$ for predicted characteristics.

\begin{figure*}[!h]
    \centering
    \includegraphics[width=\textwidth]{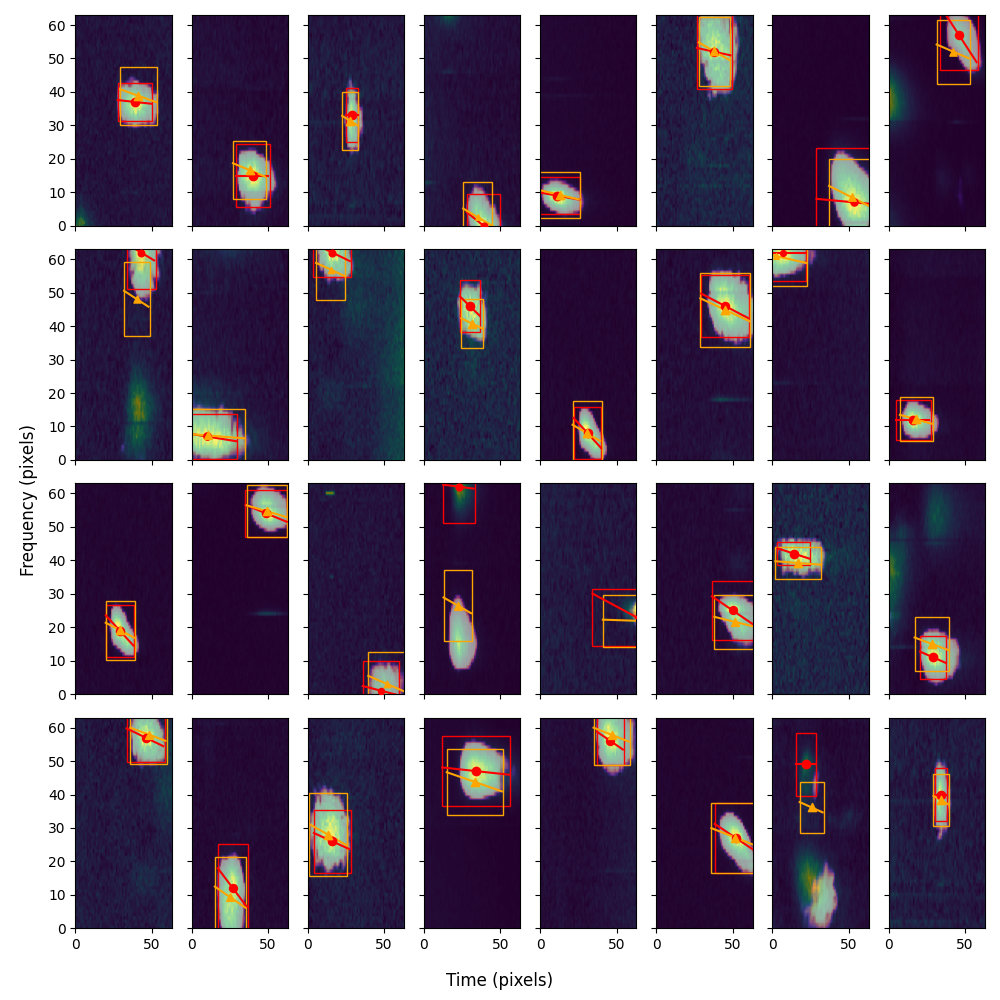}
    \caption{A comparison of the model performance compared to the ground truth. Each panel represents a sample from the validation set and depicts the dynamic spectrum of the spike burst with an overlay of the predicted segmentation mask. The ground truth spike position is given by a red circle while an orange triangle denotes the predicted position. The bounding box, determined by the spike duration and spectral width, as well as its drift rate is also plotted in red for ground truth and orange for model prediction. The axes of the panels denote time (x axis) and frequency (y axis) in pixels.}
    \label{fig:performance}
\end{figure*}

It is more interesting, and perhaps more informative, to take a detailed look at the model outputs in Figure \ref{fig:performance}.
Here we show a batch of 32 samples from the validation set and overlay the true (red) and predicted (orange) centre of the spike, a bounding box described by the total duration and spectral width, and the drift rate.
We also overlay the predicted segmentation masks in a white transparent layer.
As we expected from the histograms in Figure \ref{fig:param_hist}, the predicted characteristics mostly match well with the ground truth but not for every sample.
In particular, any time there is more than one spike in the tile, the predicted location is often somewhere between the two spikes.
This has likely occurred because the ground truth segmentation masks and spike characteristics assume only one spike is present in each tile thus the model hasn't learned how to handle multiple spikes.
We discuss possible methods to correct for this in the next section.

The predicted segmentation masks are output as polygons in the TFCat \citep{cecconi_time-frequency_2023} format.
A selection of these are currently available at \url{https://doi.org/10.25935/m5cq-f460}.
This can be queried using the Astronomical Data Query Language \citep[ADQL,][]{yasuda_astronomical_2004}.
The scripts to prepare the data and train the model are available at \url{https://gitlab.obspm.fr/pmurphy/spikenet}.
%  on the sandbox VESPA (Virtual European Solar and Planetary Access) service

\section{Discussion}
\label{discussion}

\begin{figure}[t]
    \centering
    \includegraphics[width=\columnwidth]{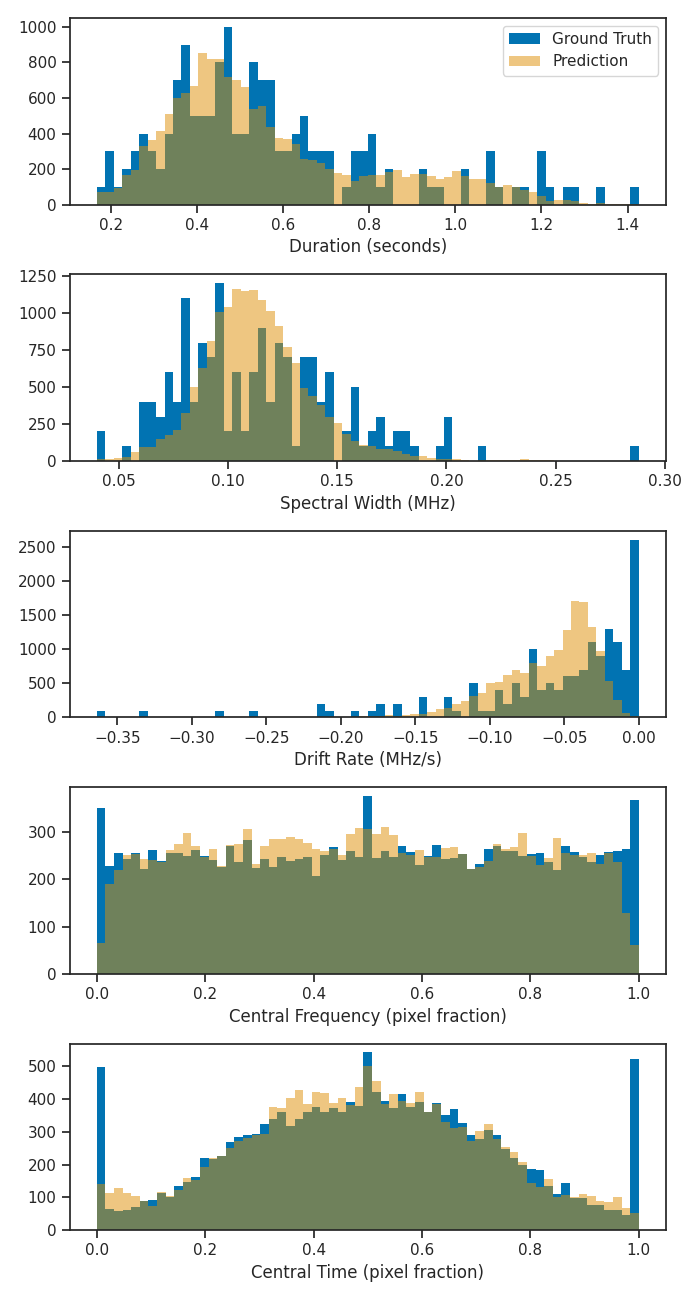}
    \caption{A comparison of the histograms for the ground truth and predicted spike characteristics. The time and frequency position are determined as a fraction of the width (height) of the dynamic spectrum in terms of pixels. We see an abundance of spikes in the validation set have centres outside the width of the tile, resulting in the peaks at 0 and 1. There is also always 1 sample with the spike directly centred in the tile which may explain the peak at 0.5 in time and frequency location.}
    \label{fig:param_hist}
\end{figure}

Machine learning methods are being applied to solar radio data in a growing number of use cases, as outlined in Section \ref{intro}.
We have found success with our CNN in segmenting and characterising solar radio spikes.
The high IoU scores obtained during training indicate an accurate recreation of the ground truth segmentation masks.
Despite the good qualitative agreement between true and predicted spike characteristics shown in Figure \ref{fig:performance}, the model still gives a mean RMS error of $\sim 23\%$.
In Figure \ref{fig:param_hist} we investigate the effect of this error on the model's predictions by comparing the histogram of each spike characteristics for the ground truth (blue) and predicted value (orange) of the validation set.
The distribution of predicted central times and frequencies are similar to the ground truths and the predicted durations follow the same overall shape as the ground truths.
The distributions for spectral width and drift rate however, do not fully agree with those of the ground truth. From this, we infer that the dominant source of error is related to the predicted spectral width. 

The histograms of the ground truth characteristics can also guide us to where the dataset selection for our model may need improvement.
For example, we see peaks in the ground truth central time and frequencies at 0, 0.5 and 1.
The peaks at 0 and 1 are likely a result of samples where the central point lies outside of the tile while the central peak is due to the validation set including a sample of every spike in the centre of the tile as well as any samples that have undergone a very small shift during augmentation described in Section \ref{method}.
The shape of the central time distribution can be attributed to the fact that the random shift applied during augmentation was based on the duration of the spike burst.
The frequency shifts, on the other hand, were a result of a uniform random distribution.
Perhaps the accuracy of the regression would improve if the time shifts were drawn from a uniform distribution as well as it will be more used to seeing spikes near the edges of a tile.

As mentioned above, the characterisation of spikes is noticeably less accurate when multiple spikes are present in a tile.
Performing semantic segmentation for distinct objects is known as instance segmentation.
One network that performs well in these kinds of tasks is called Mask R-CNN \citep[Region based CNN;][]{he_mask_2018}.
Mask R-CNN utilises a region proposal network to locate areas in each image which likely includes the desired object.
We borrow the idea of objectness from Mask R-CNN and implement it in our network as a measure of how sure the model is that it ``sees" a spike.
\cite{hou_identification_2020} have adapted the region proposal network of Faster R-CNN \citep{ren_faster_2016} to locate solar radio spikes in the \qtyrange[range-phrase=--]{150}{500}{\mega\hertz} range.
It would be a significant undertaking for the authors of this work to implement a similar network in the \qtyrange[range-phrase=--]{10}{90}{\mega\hertz} range, particularly because there is no reference in \cite{hou_identification_2020} as to where the code to train their network exists.

\subsection{Application to unseen, unlabelled data}

\begin{figure}[t]
    \centering
    \includegraphics[width=\columnwidth]{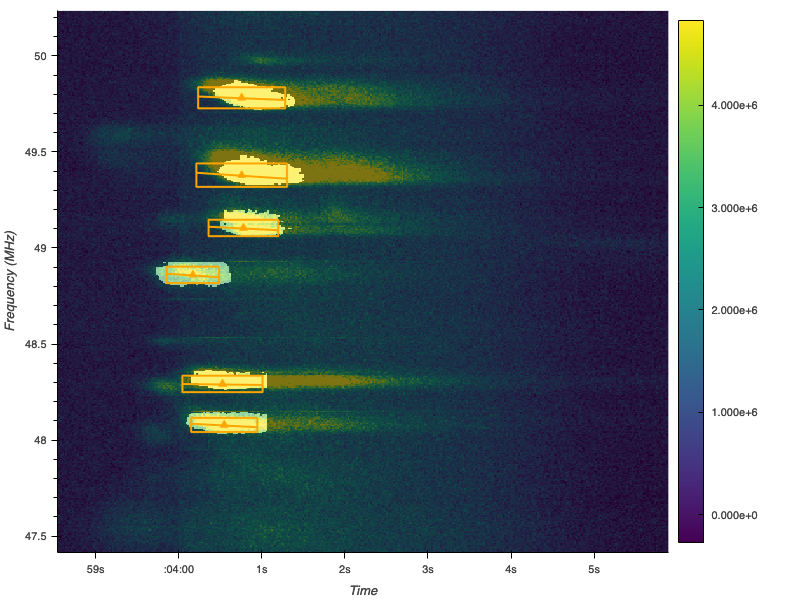}
    \caption{Performance of the model on unseen data. The panel shows a Stokes I dynamic spectrum at the same frequency and temporal resolution as Figure \ref{fig:spike-sample}. Overlayed on the dynamic spectrum is the segmentation mask detected by the CNN. Orange triangles denote the predicted centre position of the spikes while their duration, spectral width, and drift rate are shown as the orange bounding boxes and lines respectively. The segmentation masks capture the core of each spike well but fail when a spike extends across a tile.}
    \label{fig:unseen-data}
\end{figure}

The ultimate test for our machine learning model is to apply it to a dynamic spectrum that does not exist in the training, validation or test dataset and thus does not have any ground truth masks or spike characteristics.
This is instructive to qualitatively determine the robustness of the model to unseen data and inform how the model can be further developed and put into practical use.
Figure \ref{fig:unseen-data} shows a Stokes I dynamic spectrum between 11:03:59 and 11:04:05 on 2022-02-02 in the frequency range \qtyrange[range-phrase=--]{47.5}{50}{\mega\hertz} over 4 seconds.
Overlayed in transparent white are the segmentation masks as well as the central locations, marked by orange triangles, duration and spectral width, denoted by orange bounding boxes, and drift rate, denoted by orange lines.

We see that the core of the spikes are well captured by the predicted segmentation masks but the model fails to predict most of the spikes duration.
Similarly, the central point and bounding boxes do not cover the full spike.
We attribute this to the spikes not falling fully into the tiles that the original dynamic spectrum was divided into. Similar to the problem of multiple spikes appearing in one tile, this behaviour could be corrected by using a region proposal network.
Nevertheless, the performance on unseen data is promising and, if used with appropriate caution, could allow for the analysis of solar radio spikes without having to manually find them. 
% A quick analysis of the few spikes that were detected have resulted a mean drift rate and spectral width consistent with other observations \citep{Reid2021,clarkson_first_2021}

\section{Conclusion}
\label{conclusion}

We have trained a machine learning model on \num{\sim 100000} samples of solar radio spikes using an Nvidia Tesla T4 \qty{16}{\giga\byte} GPU. 
The training took 2.2 hours in total and resulted in a high $\sim 80\%$ IoU and relatively low $< 0.1$ RMS error.
The model successfully produced segmentation masks for radio spikes in the validation set and predicted the spike characteristics, i.e. location in time and frequency, duration, spectral width and drift rate.
The model does have some shortcomings when applied to unseen data.
For example, a certain number of spike masks are abruptly cut off and the model was less successful in determining the position of spikes when more than one spike appeared in the $64 \times 64$ pixel dynamic spectrum sample. 
This is due to the original dynamic spectrum being divided into $64 \times 64$ tiles thus, any burst not fully in one tile is unlikely to be fully recovered in the mask. 
A region proposal network to first determine where spikes are likely to be is the best solution for this.

With the current data rate generation of LOFAR and NenuFAR, not to mention the ongoing development of next generation low frequency arrays such as the Square Kilometre Array (SKA) which are set to produce petabytes of data, the ability to automatically detect fine scale features in solar radio observations has become increasingly important.
This task is ideally suited to machine learning methods, the latest of which we have presented here.
Our CNN drastically reduces the time to locate and identify solar radio spikes from minutes to less than a second per spike. While the accuracy of the predicted spike characteristics could be improved, the use of our CNN trivialises finding spikes hidden in solar radio data and allows for more time to be dedicated to their analysis using whichever methods a particular researcher desires. 
Future development of SpikeNet could see it used in near real-time on NenuFAR solar observations to dynamically determine an appropriate time resolution. For example, should SpikeNet detect a certain number of spikes over a given time period, keep this data at its native resolution. Otherwise, the time resolution should be rebinned. This will alleviate pressure on the fixed storage space allocated to solar observations with NenuFAR. The expected data rates of the SKA \citep{Scaife2020} mean that dynamically adapting the recording resolution could drastically reduce the storage required for solar observations in the future.

\section*{Acknowledgements}
This work has been supported by the Europlanet-2024 Research Infrastructure project, which received funding from the European Union's Horizon 2020 research and innovation programme under grant agreement No 871149 and a grant from Paris Observatory science council under the MINERVA project.
We acknowledge the use of the Nançay Data Center (CDN – Centre de Données de Nançay) facility. The CDN is hosted by the Observatoire Radioastronomique de Nançay (ORN) in partnership with the Observatoire de Paris, the Université d’Orléans, the Observatoire des Sciences de l’Univers d’Orléans (OSUC) and the French Centre National de la Recherche Scientifique (CNRS). The CDN is supported by the Région Centre-Val de Loire (département du Cher). The ORN is operated by the Observatoire de Paris, associated with the CNRS.
The research leading to these results has received funding from the European Union's Horizon Europe Programme under the EXTRACT Project, grant agreement n° 101093110. 

\newblock
\bibliographystyle{aa} 
\bibliography{references1.bib}
% \newblock
\onecolumngrid \footnotesize
\end{document}